\documentclass[12pt,dvips]{article}
\usepackage{epsfig}

\begin{document}
\title{
\vspace{-5.0cm}
\begin{flushright}
{\normalsize UNIGRAZ-}\\
\vspace{-0.3cm}
{\normalsize UTP-}\\
\vspace{-0.3cm}
{\normalsize 29-05-96}\\
\end{flushright}
\vspace*{1cm}
\vfill
Non-Gaussian fixed point in four-dimensional pure compact U(1) gauge
theory on the lattice}

\author
{\bf J.~Jers{\'a}k \\  \\
Institut f{\"u}r
Theoretische Physik E,\\RWTH Aachen, Germany \\ \\
\bf C.~B.~Lang\\ \\
Institut f{\"u}r Theoretische Physik, \\
Karl-Franzens-Universit\"at Graz, Austria \\ \\
\bf T.~Neuhaus\\ \\
FB8 Physik, BUGH Wuppertal, Germany}
\date{\today}
\maketitle
\thispagestyle{empty}

\begin{abstract}
The line of phase transitions, separating the confinement and the
Coulomb phases in the four-dimensional pure compact U(1) gauge theory
with extended Wilson action, is reconsidered. We present new numerical
evidence that a part of this line, including the original Wilson
action, is of second order. By means of a high precision simulation on
homogeneous lattices on a sphere we find that along this line the
scaling behavior is determined by one fixed point with distinctly
non-Gaussian critical exponent $\nu = 0.365(8)$. This makes the
existence of a nontrivial and nonasymptotically free four-dimensional
pure U(1) gauge theory in the continuum very probable. The universality
and duality arguments suggest that this conclusion holds also for the
monopole loop gas, for the noncompact abelian Higgs model at large
negative squared bare mass, and for the corresponding effective string
theory.
\end{abstract}

\section{Introduction}

In contradistinction to lower dimensions, the only firmly established
quantum field theories in four dimensions (4D) are either
asymptotically free or so-called trivial theories. Both are defined in
the vicinity of Gaussian fixed points, i.e. of noninteracting limit
cases. In spite of numerous suggestions and circumstantial evidence,
until now no candidate for a non-Gaussian fixed point in 4D has been
established.

The suitability of numerical simulations on the lattice for a
confirmation of the existence of non-Gaussian fixed points, and for an
investigation of their properties, has been demonstrated in dimensions
lower than four in numerous applications. For example, non-Gaussian
values of critical exponents can be determined by means of finite-size
scaling (FSS) or renormalization group (RG) analysis.  Several attempts
to use a similar approach in 4D have encountered severe problems,
however.

In this letter we reconsider the oldest candidate for a non-Gaussian
fixed point in the 4D lattice field theory, the phase transition
between the confinement and the Coulomb phases in the pure compact U(1)
gauge theory \cite{Wi74,Gu80} with Wilson action and extended Wilson
action.  After various pioneering studies, e.g.
\cite{CrJa79,LaNa80,Bh82}, the more detailed investigations were
hindered mainly by a weak two-state signal \cite{JeNe83,EvJe85}.  This
obscures the order of the phase transition and makes it difficult to
determine critical exponents reliably. We demonstrate that the problems
encountered, when considering the continuum limit at this phase
transition, can be surmounted. The clues are the observation that the
two-state signal disappears on lattices with spherelike topology, the
construction of homogeneous spherical lattices, the use of modern FSS
analysis techniques, and larger computer resources.

We find that at the confinement -- Coulomb phase transition at strong
bare gauge coupling, $g$ = $O(1)$, the model exhibits a $2^{\rm nd}$
order scaling behavior well described by the values of the correlation
length critical exponent $\nu$ in the range $\nu = 0.35 - 0.40$. The
measurements have been performed at various couplings and by different
methods. The most reliable determination gives
\begin{equation}
         \label{NU}       \nu = 0.365(8)  .
\end{equation}
These results are quite different from $\nu=0.25,$ expected at a
$1^{\rm st}$ order transition, as well as from $\nu = 0.5,$ obtained
in a Gaussian theory or in the mean field approximation. This
strongly suggests the existence of a continuum pure U(1) gauge theory
with properties different from theories governed by Gaussian fixed
points with or without logarithmic corrections. It can be obtained
from the lattice theory by the RG techniques.

Detailed numerical evidence for these claims will be presented
elsewhere\cite{JeLa96b}.  Some preliminary results have been published
in Refs.~\cite{JeLa95,HoJa96}.

Since the pure U(1) lattice gauge theory with the Villain (periodic
Gaussian) action presumably belongs to the same universality class
\cite{LaPe96}, rigorous dual relationships imply that also the
following 4D models possess a continuum limit described by the same
fixed point: the Coulomb gas of monopole loops\cite{BaMy77}, the
noncompact U(1) Higgs model at large negative squared bare mass
(frozen 4D superconductor) \cite{Pe78,FrMa86}, and an effective string
theory equivalent to this Higgs model\cite{PoSt91}.

These findings raise once again the question, whether in strongly
interacting 4D gauge field theories further non-Gaussian fixed points
exist, that might be of use for theories beyond the standard model.

\section{Earlier results}

The pure compact U(1) gauge theory on the 4D cubic lattice with
periodic boundary conditions (4D torus) can be described by the
extended Wilson action \cite{Bh82}
\begin{equation}\label{ACTION}
         S = -\sum_P w_P
              \left [\beta \cos(\Theta_P) + \gamma
                \cos(2\Theta_P)\right ],
\end{equation}
with $w_P = 1$. Here $\Theta_P \in [0,2\pi)$ is the plaquette angle,
i.e. the argument of the product of U(1) link variables along a
plaquette $P$. Taking $\Theta_P = a^2gF_{\mu\nu}$, where $a$ is the
lattice spacing, and $\beta + 4\gamma = 1/g^2$, one obtains for weak
coupling $g$ the usual continuum action $S =\frac{1}{4} \int
d^4xF_{\mu\nu}^2$.

In one of the very first studies \cite{LaNa80}, restricted to the
$\gamma = 0$ case (Wilson action\cite{Wi74}) and small lattices, a
behavior consistent with a $2^{\rm nd}$ order phase transition at
$\beta \simeq 1$ was observed, and $\nu \simeq 1/3$ was found.
However, any inference about the continuum limit has been hindered by
the subsequent observation of a two-state signal on larger, but finite
lattices\cite{JeNe83}. This could either be a finite-size effect, or
it could imply that the phase transition at $\gamma = 0$ is actually
of $1^{\rm st}$ order, preventing a continuum limit at $\gamma = 0$.

In the model with the extended Wilson action (\ref{ACTION}), it was
found that the confinement-Coulomb phase transition is clearly of
$1^{\rm st}$ order for $\gamma \ge 0.2$, and weakens with decreasing
$\gamma$ \cite{Bh82,EvJe85}. Various studies suggested that the
transition becomes $2^{\rm nd}$ order at slightly negative
$\gamma$ \cite{EvJe85}, or around $\gamma = 0$\cite{GuNo86,La86}.

The order of the transition at $\gamma = 0$ has remained a
controversial subject \cite{LaNe94a,KeRe94,BaFo94,LiBo95}. Though the values
$\nu \simeq 0.3 - 0.4$ have been obtained consistently by various
methods \cite{LaNa80,Bh82,GuNo86,La86,JeNe85,LaRe87}, the continuum
limit has not appeared to be possible.

Also the hope that the continuum limit might be taken at least at
negative $\gamma$ was spoiled by the observation of a
weaker, but still significant, two-state signal on finite lattices
even there\cite{EvJe85}. Though this signal is probably only a
finite-size effect, and the transition in the infinite volume limit is
genuinely of $2^{\rm nd}$ order, it impedes a precise FSS and RG
analysis.

It was known that monopole loops winding around the toroidal lattice
occur \cite{GrJa85,GuNo86} and cause difficulties in simulations with
local update algorithms. Suspecting that this might be a reason for
the two-state signal, two of the present authors performed simulations
at $\gamma = 0$, using the 4D surface of a 5D cubic lattice instead of
the torus. They observed that on such lattices with spherelike
topology the two-state signal vanishes \cite{LaNe94a}. This
suggests that the two-state signal at $\gamma \le 0$ is related to the
nontrivial topology of the toroidal lattice.

Related observations have been made for the Schwinger model
\cite{GaLa95}.  On the other hand, it has been checked in spin models,
that weak two-state signals are not washed out on lattices with
spherelike topology, if they are due to a genuine $1^{\rm st}$ order
transition\cite{HoJa96}.

However, the lattice on the surface of a cube is rather inhomogeneous
and causes complex finite-size effects, preventing a reliable FSS
analysis.

\section{Present methods}

For our present study at $\gamma \le 0$, we have chosen again lattices with
spherelike topology. To alleviate the problem of inhomogeneity, we
have used lattices obtained by projecting the 4D surface
SH[N] of a 5D cubic lattice $N^5$ onto a concentric 4D sphere.
On such a spherical lattice S[N], the curvatures concentrated on
the corners, edges, etc., of the original lattice SH[N] are
approximately homogenized over the whole sphere by the weight factors
\begin{equation}
             w_P = A'_P/A_P  \label{WP}
\end{equation}
in the action (\ref{ACTION}). $A_P$ and $A'_P$ are the areas of the
plaquette $P$ on S[N], and of its dual, as on any irregular, e.g. random
lattice \cite{ChFr82}.  These areas are determined by means of a
two-triangle approximation of the plaquettes.

It has been checked in some spin and gauge models with $2^{\rm nd}$
order transitions that universality for spherical lattices holds, and
that the FSS analysis works very well if
$V^{1/D}$ is used as a linear size parameter,
$V\equiv\frac{1}{6}\sum_P w_P $ being the volume of the sphere
S[N] \cite{HoJa96,HoLa96}.

Another new feature in our study of the theory is the FSS analysis
of the first zero $z_0$ of the partition function in the complex plane
of the coupling $\beta$ (Fisher zero). Applying the multihistogram
reweighting method \cite{FeSw89a} for its determination, the expected
FSS behavior
\begin{equation}
          \mbox{Im}\,z_0 \propto V^{-1/D\nu}    \label{IMZ}
\end{equation}
has been used for measuring $\nu$. This has turned out to be superior
to --- though consistent with --- the more common FSS analysis of
specific-heat and cumulant extrema.

Finally, we have performed the FSS study of the confinement-Coulomb transition
not only at $\gamma = 0$, but also at $\gamma = -0.2$ and $-0.5$. This
allows us to compare the behavior of the system at $\gamma = 0$, where
the order of the transition is disputed, with the commonly expected
$2^{\rm nd}$ order behavior at negative $\gamma$, and to test the
universality of the critical properties.

\section{Results}

We have performed simulations \cite{JeLa96b} on S[N] for $N$
between 4 and 12.  Note that $S\,[12]$ has about $(19.6)^4$ lattice points.
The values of $\beta$ were chosen in the immediate neighborhood of its
critical values for $\gamma = 0, -0.2$, and $-0.5$. For each lattice
size at each $\gamma$ value, we have accumulated typically $10^6$
updates distributed over 8 -- 12 $\beta$ points.

\begin{figure}[htp]
\begin{center}
    \epsfig{file=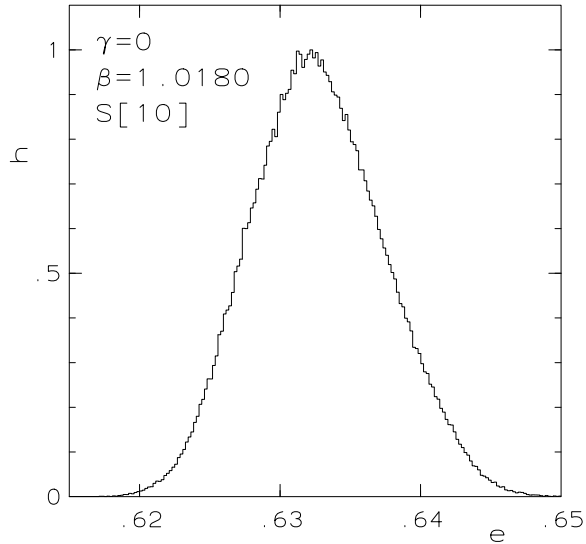,width=8cm}
\caption{Histogram of the energy density $e$ on the $S\,[10]$ lattice
(about $16^4$ points) at $\gamma = 0$ and $\beta  = 1.0180$, very close
to the pseudocritical point $\beta_{pc} = 1.01835(4)$ on that lattice.
$h$ is the relative occurence of $e$-values. }
\label{fig:1}
\end{center}
\end{figure}

We have found no indication of a two-state signal, neither in the
individual, nor in the multihistogram distributions of $e = (\sum_P
w_P \cos (\Theta_P))/(\sum_P w_P)$, at any of the three $\gamma$
values. This is demonstrated in Fig.~1 for $\gamma = 0$.  The values
of the studied cumulants at their respective extrema are compatible
with a $2^{\rm nd}$ order transition.  The critical behavior at all
three $\gamma$ values is very similar, except that the transition
weakens with decreasing $\gamma$, which means that larger lattices are
needed for the same height of the specific-heat peak.

In test runs at $\gamma = +0.2,$ a two-state signal has been clearly
observed. This confirms that at sufficiently large positive $\gamma$
the phase transition is of $1^{\rm st}$ order, and that the spherical
lattice S[N] does not wash out such a signal.

Furthermore, at all three investigated values $\gamma \le 0$, the FSS
analysis assuming a $2^{\rm nd}$ order transition works remarkably
well and leads to consistent results for all observables \cite{JeLa96b}.  All our
evidence thus points towards the conclusion that the phase transition is
of 2$^{\rm nd}$ order for $\gamma \le 0$.


\begin{table}[htb]
\caption{Results for $\nu$ from fits to $\mbox{Im}z_0$, $c_{V,max}$
and $V_{CLB,min}$ at various $\gamma$. The indicated errors are 
statistical.}
\label{tabfit}
\begin{center}
\begin{tabular}{r|ccc}
\hline
$\gamma$ &  $\mbox{Im}\,z_0$ & $c_{V,max}$ & $V_{CLB,min}$ \\
\hline
\hline
0.    & 0.345(3)   & 0.361(6)   & 0.361(6)  \\
-0.2  & 0.378(7)   & 0.374(6)   & 0.365(6)  \\
-0.5  & 0.368(8)   & 0.404(9)   & 0.396(9) \\
\hline
\end{tabular}
\end{center}
\end{table}


In Table~1 we present results for the critical exponent $\nu$ obtained
from all the data with $N \ge 4$.  The most reliable ones come from
the FSS analysis of the Fisher zero (first column).  The approximate
agreement between the obtained values of $\nu$ at all three $\gamma$
values demonstrates that the confinement-Coulomb critical line belongs
to one universality class. The FSS behavior of $\mbox{Im}\,z_0$
according to (\ref{IMZ}), and the consistency of this behavior at
different $\gamma$ are illustrated in Fig.~2. The value (\ref{NU}) has
been obtained from a joint fit by means of (\ref{IMZ}) to the data for
$\mbox{Im}\,z_0$ with $N \ge 6$ (parallel straight lines in Fig.~2).

The next two columns in Table~1 show the results from the FSS analysis
of the maximum of the specific-heat $c_{V,max}$ and the minimum of the
Challa-Landau-Binder cumulant $V_{CLB,min}$. Here certain assumptions
about nonleading terms have been necessary \cite{JeLa96b}, and small
systematic errors are therefore possible. Nevertheless, all the shown
results are consistent with $\nu$ lying in the interval $0.35 - 0.40$,
thus supporting universality.

\begin{figure}[htbp]
\begin{center}
     \epsfig{file=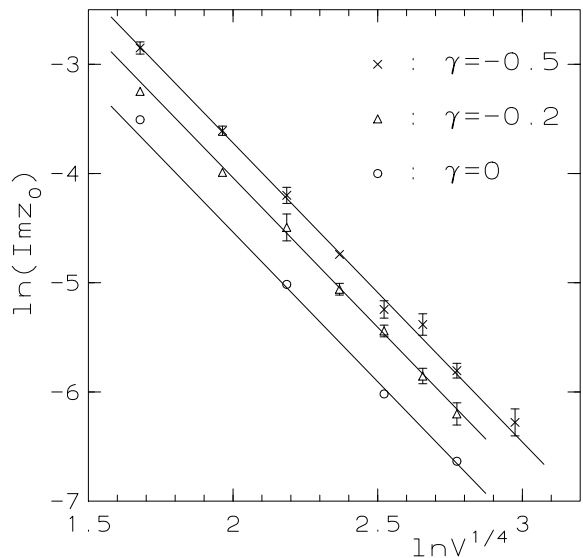,width=8cm}
\caption{Joint fit to $\mbox{Im}z_0$ according to (\protect\ref{IMZ})
for all three $\gamma$ values and $N \ge 6$. The values of $L =
V^{1/4}$ correspond to $N = 4-10,12$.} \label{fig:2}
\end{center}
\end{figure}

\section{Discussion}

The physical content of the continuum limit of the pure compact U(1)
gauge theory at the confinement-Coulomb phase transition, discussed
e.g. in \cite{Pe78,Cr81}, depends on the phase from which the critical
line is approached. In the confinement phase, a confining theory with
monopole condensate is expected, as the string tension scales with a
critical exponent consistent with the value (\ref{NU})\cite{JeNe85}.
The physical spectrum consists of various gauge balls. In the Coulomb
phase, massless photon and massive magnetic monopoles, both already
observed in Monte Carlo simulations \cite{BePa84,PoWi91}, should be
present. The renormalized electric charge $e_r$ is large but
finite\cite{Ca80,JeNe85}, and has presumably a universal
value\cite{Ca80,Lu82,JeNe85}. The numerical result $e_r^2/4\pi =
0.20(2)$\cite{JeNe85} agrees with the L\"uscher bound \cite{Lu90}, as
explained in \cite{PoWi91}.

To our knowledge, the existence of such continuum quantum field
theories in 4D is in no way indicated by the perturbation expansion.
The non-Gaussian character of the fixed point might be rather
understood as a consequence of the complex dynamics of the systems
obtained in the dual representation of the theory with Villain action
\cite{BaMy77,Pe78,FrMa86,PoSt91,PoWi91}.

There are some questions deserving further discussion. For $\gamma <
0$, the studied theory does not satisfy reflection positivity, which
is a sufficient, albeit not necessary condition for unitarity. If the
phase transition in the reflection positive case at $\gamma = 0$ is of
2$^{\rm nd}$ order, as strongly suggested by our results, then
unitarity should hold also for $\gamma < 0$ by universality arguments.
If it is of weak $1^{\rm st}$ order, unitarity at $\gamma < 0$ is made
plausible by our finding that the scaling behavior at $\gamma = 0$ (on
lattices of limited size), and at $\gamma < 0$ is the same, and that
the regions with $\gamma < 0$ and $\gamma \ge 0$ are connected by the
RG flows\cite{La86}.

Though improbable, it is not completely excluded that the phase
transition we have studied is of $1^{\rm st}$ order for any $\gamma \le
0$ \cite{Ha88a}, but so weak, that the two-state signal cannot be
detected by the currently available numerical means. Then the correlation
length would remain finite for all finite $\gamma$. Presumably, the
effective cutoff, arising in this way, may be made arbitrarily large by
suitably decreasing $\gamma$. However, unitarity then might be
questionable.

At small $\gamma > 0$, where the order of the transition probably
changes, a tricritical point with special values of indices is
expected\cite{EvJe85}. As the domain of dominance of such a point is
unknown, it could be that the measured value of $\nu$, and the corresponding
non-Gaussian fixed point are actually tricritical. Alternatively,
the change of the order might be more complicated than
in metamagnets with tricritical points, since in our case
only nonlocal order parameters are available.

Though this may seem plausible, we cannot firmly conclude that the
two-state signal observed on toroidal lattices is due to the monopole
loops winding around them.  The evidence for such an explanation is
controversial \cite{GrJa85,KeRe94,BaFo94,LiBo95}, and the question,
which configurations make the difference between the compact U(1)
gauge theory on finite toroidal and spherical lattices, requires
further study. The thermodynamic limit ought to be the same.

We thank D.P.~Landau, T.~Lippert, M.I.~Polikarpov, and
U.-J.~Wiese for discussions. The computations have been performed on
the CRAY-YMP of HLRZ J\"ulich and at the Parallel Compute Server of
KFU Graz. J.J. was supported by Deutsches BMBF.


\end{document}